\documentstyle[psfig]{lamuphys}
\makeatletter
\let\chapter\hid@chapter
\makeatother
\begin{document}
\pagenumbering{arabic}
\title{Radio Continuum and Emission Line Morphologies of Southern Seyfert 
Galaxies}

\author{Z.\,Tsvetanov\inst{1}, 
	R.\,Morganti\inst{2,3}, 
	R.A.E.\,Fosbury\inst{4}, 
	M.G.\,Allen\inst{5}, and 
	J.\,Gallimore\inst{6}}

\institute{Johns Hopkins University, Baltimore, MD 21218, USA
\and
Australia Telescope National Facility, PO Box 76, Epping NSW 2121, Australia
\and
Instituto di Radioastronomia, CNR, I-40129 Bologna, Italy
\and
ST-ECF, D-85748 Garching bei M\"unchen, Germany
\and
Mount Stromlo and Siding Spring Observatories, ACT 2611, Australia
\and 
Max Planck Institut f\"ur Extraterrestische Physik, Garching bei M\"unchen, 
Germany}

\maketitle

\vspace{-3mm}
\section{Background}
\vspace{-2mm}
Modern active galactic nuclei (AGN) research is greatly concerned with
geometry. The radiation and absorption anisotropies, and the
orientation to the line of sight, are fundamentally important for both
the object classification and for understanding the physical nature of
the copious energy release.

There are two types of observed anisotropies -- highly collimated
radio emission, i.e., beams, jets (opening angle $\theta \sim
10^{\circ}$), and ionization cones ($\theta \sim100^{\circ}$). The
axial symmetry is determined on a subparsec scale but the two types of
collimated radiation are traced out to kiloparsec scales. It is
particularly important to understand the connection between the AGN
axis and any global symmetry properties of the host galaxy. With no
exception, the radio and ionization cone axis are aligned to within the
measurement errors (e.g., Wilson \& Tsvetanov 1994), but there is
little relation to other galaxian scale axes.

To address the important questions of the AGN -- host galaxy
relationship we have collected extensive optical emission line and
radio continuum imaging data for a volume limited sample of southern
Seyfert galaxies.  Our sample consists of 50 well classified galaxies
with $cz \leq 3600$ km s$^{-1}$ and $\delta \leq 0^{\circ}$.

\vspace{-3mm}
\section{Observations}
\vspace{-2mm} Optical [{\sc O iii}] $\lambda$5007 and H$\alpha$+[{\sc
N ii}] emission line and their adjacent continua images of all the
galaxies in the sample were obtained at ESO using the 2.2 m, NTT and
3.6 m telescopes.  The typical resolution of the emission line maps is
1$''$, with a noise level of $\sim1\times10^{-16}$ ergs cm$^{-2}$
s$^{-1}$ arcsec$^{-2}$. In addition to the emission line maps, for
each galaxy in the sample we have formed an excitation/reddening map,
the ratio [{\sc O iii}] $\lambda$5007 / (H$\alpha$+[{\sc N ii}]), and
a continuum color map.  The later is affected by a combination of
extinction and color effects.

New radio continuum observations were obtained for 29 of the galaxies
in the sample. Objects with $-30^{\circ} < \delta < 0^{\circ}$, 8 in
total, were observed with the Very Large Array (VLA) at 4.9 GHz (6 cm)
and objects with $\delta < -30^{\circ}$ were observed with the
Australia Telescope Compact Array (ATCA) at 8.4 GHz (3 cm). Both the
VLA and ATCA radio maps have a resolution of $\sim1''$ matching that
of the optical images.  All, but one of the observed sources were
detected above the noise limit of $\sim0.15$ mJy. Our radio
observations were combined with data available from the literature to
achieve almost 85\% coverage of the sample.

\vspace{-2mm}
\section{Highlights of Results}
\vspace{-2mm} 
In the radio, 30\% of the sources show linear structure, 25\% are only
slightly resolved or diffused, and 45\% remain unresolved at the
$\sim1''$ resolution. As in previous work, a correlation is found
between the size of the radio structure and the radio power. The radio
sources in Seyfert 2 galaxies have, on average, larger linear size
than their type 1 counterparts (see Fig.\ 1), but there is no
significant difference in radio power between types 1 and 2, although
all the most powerful objects appear to be Seyfert 2's.  No
significant difference is found in the spectral indices of the two
Seyfert types.

Extended emission is common in Seyfert galaxies -- essentially all
objects observed show extended H$\alpha$+[{\sc N ii}] and nearly 50\%
show extended [{\sc O iii}] emission. At least 40\% (18 out of 50) of
the galaxies show high excitation extended emission well outlined in
the excitation map. The morphology of the high excitation extended
emission line region vary from linear to conical to S-shaped and even
X-shaped. Almost exclusively all elongated EELR are in Seyfert 2
galaxies. The orientation of the EELR appears to be random relative to
either the major (or minor) axis or relative to the non-axisymmetric
structures, such as bars or ovals, when present, and there is a hint 
of relation to the morphological type (Fig.\ 2).

\begin{figure}[ht]
%\vspace{5.5cm}
\vspace{-7mm}
\centerline{\psfig{figure=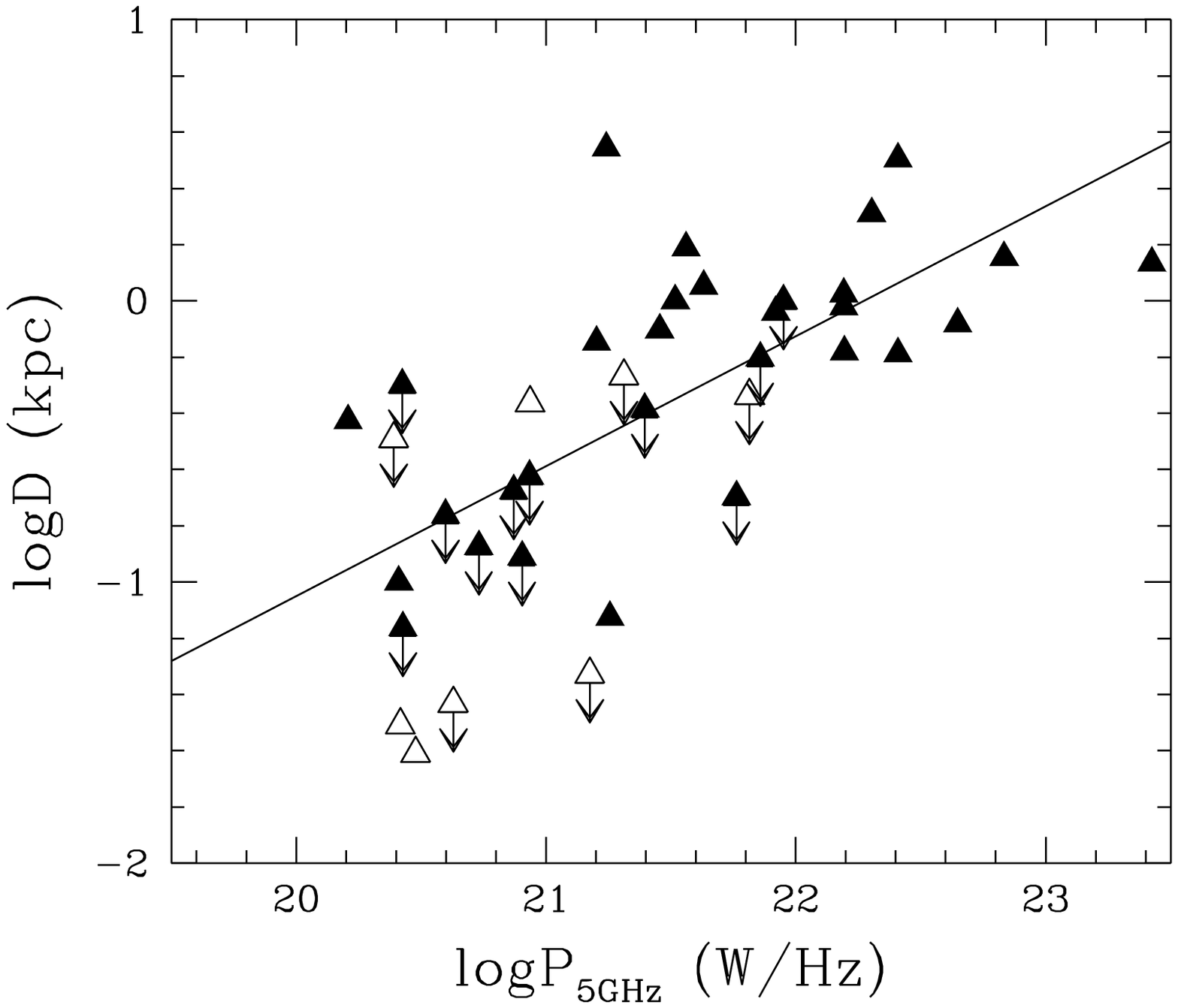,width=6cm}
            \psfig{figure=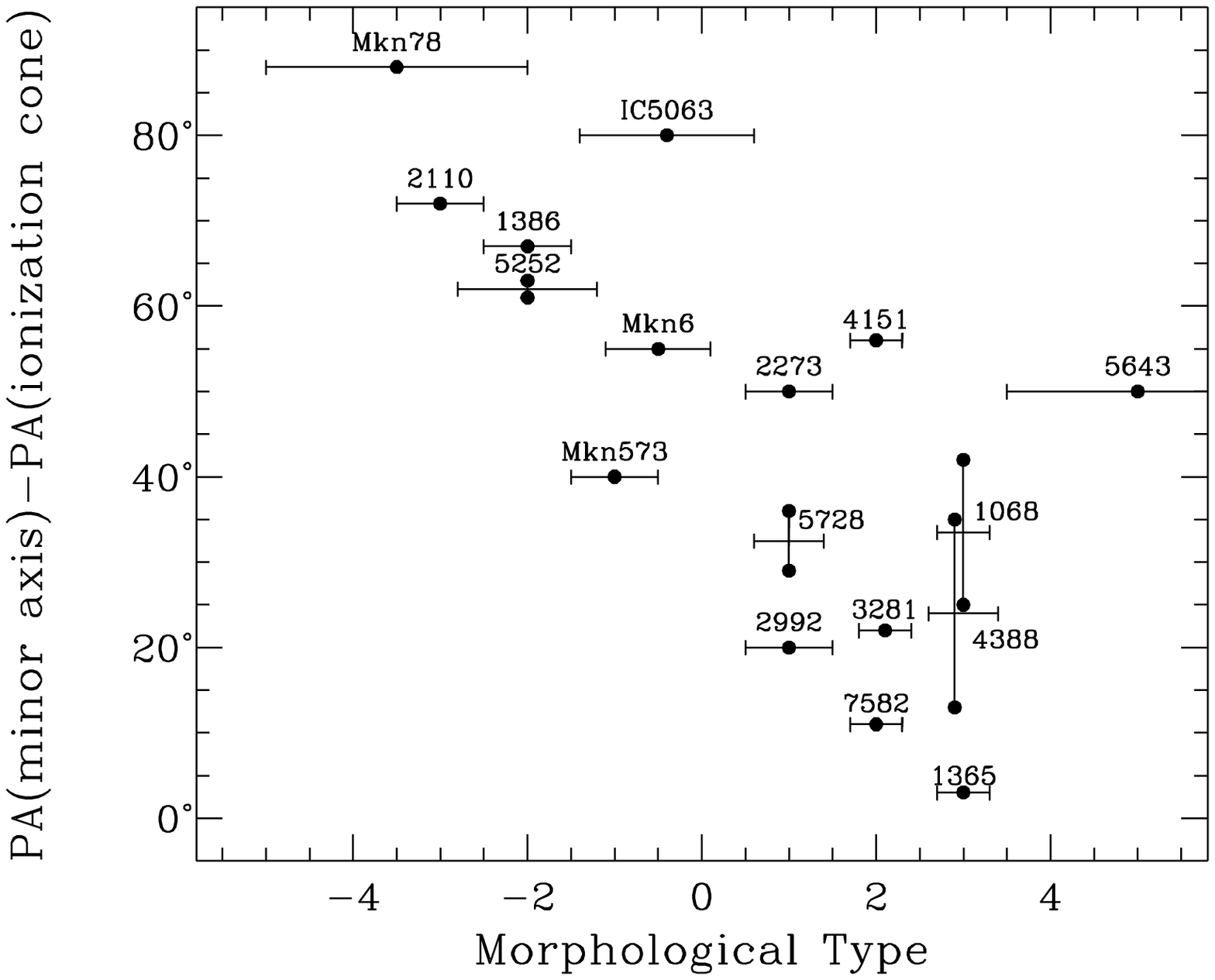,width=6cm}}
\vspace{-17mm}
\end{figure} 

\noindent\parbox[t]{57mm}{ {\small\bf Fig.1.} \small Radio power
versus linear size of the radio structure. Open triangles are Seyfert
1's, filled symbols are Seyfert 2's. The K-S test suggests that the
two distributions are significantly different. The median size of the
radio structure is 0.32 kpc in Seyfert 1's and 0.66 kpc in Seyfert 2's
(56\% and 44\% of measurements being upper limits, respectively).}
{\hspace{5mm}}
\parbox[t]{57mm}{ {\small \bf Fig.2.} \small Orientation of the
galactic axis with respect to the ionization cone axis plotted as a
function of the morphological type of the host galaxy. Objects are
indicated by their NGC or Markarian number.  Vertical lines connect
the values on opposite sides of the nucleus. There seems to exist a
clear tendency, but selection effects may play significant role.}

\end{document}